\input phyzzx
\hsize=417pt %\vsize=600pt \baselineskip=20pt \maxdepth=0.2pt
\sequentialequations
\Pubnum={ EDO-EP-7}
\date={ \hfill October 1996}
\titlepage		
\vskip 32pt
\title{ Black Hole Radiation inside Apparent Horizon in Quantum 
Gravity }
\author{ Akio Hosoya \footnote\dag {E-mail address: 
ahosoya@th.phys.titech.ac.jp}}
\vskip 12pt
\address{ Department of Physics, Tokyo Institute of Technology,
Oh-Okayama, Meguro-ku, Tokyo 152, JAPAN }                          
\vskip 15pt
\centerline{and}
\vskip 12pt
\author{Ichiro Oda \footnote\ddag {E-mail address: 
sjk13904@mgw.shijokyo.or.jp}}
\vskip 12pt
\address{ Edogawa University,                                
          474 Komaki, Nagareyama City,                        
          Chiba 270-01, JAPAN     }                          
%          
%
%              The titlepage ends at this place.
%
%
%======================================================================%
%
\abstract{ We study a black hole radiation inside the apparent horizon 
in quantum gravity. First we perform a canonical quantization for 
spherically symmetric geometry where one of the spatial 
coordinates is dealt as the time variable since we would like to consider 
the interior region of a black hole. Next this rather general formalism is 
applied for a specific model where the ingoing Vaidya metric is 
used as a simple model of an evaporating black hole. Following 
Tomimatsu's idea, we will solve analytically the Wheeler-DeWitt equation 
in the vicinity of the apparent horizon and see that mass-loss rate of a black 
hole by thermal radiation is equal to the result obtained by Hawking in 
his semiclassical treatment. The present formalism may have 
a wide application in quantum gravity inside the horizon of a black hole 
such as mass inflation and strong cosmic censorship etc. } 
\endpage
%
%=========================================================================%
%
%        Macros
%

\def\sp(#1){\noalign{\vskip #1pt}}

%
%
%
%=========================================================================%
%
%   This part is the meat of body.
%
\topskip 30pt
\par
\leftline{\bf 1. Introduction}	
\par
The canonical formalism of general relativity was pioneered in the 
sixties by Dirac [1], and by Arnowitt, Deser and Misner (ADM) [2,3]. 
Afterward an extension of the canonical formalism to systems with a black 
hole was undertaken by Hajicek where special attention was paid for the 
properties of the apparent horizon and the choice of hypersurfaces 
covering the spacetime between the apparent horizon and the spatial 
infinity [4,5]. Recently Kucha$\hat r$ has given a detailed analysis of the 
geometrodynamics of the Kruskal extention of the Schwarzschild black hole 
[6]. 

In this article, we would like to study the canonical formalism of a 
system with a spherically symmetric black hole only in the interior 
region bounded by the apparent horizon and the singularity. In this 
region the Killing vector field $\partial \over {\partial t}$ becomes 
spacelike, while it does timelike in the exterior region. 
Consequently one must foliate the interior region of a black hole by a 
family of spacelike hypersurfaces, for example, $r = const$. One will see 
that it is straightforward to extend Hajicek's works [5] performed 
in the exterior region to the present case. 

So far, the region inside the event horizon of a black hole has been 
received little attention because it is physically of no relevance for 
external observers outside the horizon. However, this 
situation has remarkably changed by the recent development of 
understanding the internal geometry near the Cauchy horizon inside the 
Reissner-Nordstr$\ddot o$m black hole, what we call, the mass inflation [7,8,9], 
and the phenomenon of the smearing of black hole singularity in quantum 
gravity [10,11]. Anyway since both spacetime singularity and the Cauchy 
horizon exhibit highly pathological behavior in the classical theory of 
general relativity, and indeed it is expected that quantum effects play a 
dominant role around them, the studies of physics inside the horizon 
of a black hole may give us some important clues for constructing a 
theory of quantum gravity in future. From this viewpoint, in this 
article we will present a the canonical formalism which describes the 
region inside the horizon of a 
spherically symmetric black hole. As one of 
applications of this canonical formalism, we shall consider the black 
hole radiation [12]. The analysis follows the idea by Tomimatsu [13] that
the Hamiltonian and supermomentum constraints have a simple and tractable
form, and consequently it is possible to solve the Wheeler-DeWitt 
equation near the apparent horizon. The mass-loss rate by the black hole 
radiation will be shown to be equal to that evaluated by Hawking in 
the semiclassical 
approximation. Hawking's discovery of thermal radiation from quantum 
black holes was a pivotal event in the development of a quantum theory of 
gravity [14], thus the results obtained in this paper seems to encourage 
us to investigate various related problems further along the present 
canonical formalism. 

The article is organized as follows. In section 2, we consider a system 
of fields in a four dimensional spacetime containing gravity, and 
transform it to the ADM first-order form. In section 3, we apply the 
canonical formalism constructed in section 2 for a calculation of the 
mass-loss rate by the black hole radiation. The last section is devoted to 
conclusion.

\vskip 12pt
\leftline{\bf 2. Canonical formalism}	
\par
We begin by constructing a canonical formalism of a spherically symmetric 
system with a black hole. In this paper, since we have a mind to apply 
the canonical formalism for various physically interesting phenomena 
occurring inside the horizon of a black hole we will construct a 
canonical formalism which holds only in the internal region between the 
apparant horizon and the singularity of a black hole. First of all, 
following the conventional procedure, one needs to choose arbitrary 
spherically symmetric spacelike hypersurfaces to form a foliated 
structure of the spacetime. The important point is that the radial 
coordinate plays a role of time inside the horizon in the spherically 
symmetric coordinate system. As a simple choice, let us take a choice of 
$x^1 = const$ hypersurfaces to slice the spacetime. Later we will take 
the simplest choice $x^1 = r$.  

The four dimensional action which we consider in this section is of the form
%%%%%%%%%%%%%%%%%%%%%%%%%%%%Equation%%%%%%%%%%%%%%%%%%%%%%%%%%%%%%%%%%%
$$ \eqalign{ \sp(2.0)
S = \int \ d^4 x \sqrt{-^{(4)}g} \bigl[ {1 \over 16 \pi} {}^{(4)}R - {1 \over 
4 \pi} {}^{(4)}g^{\mu\nu} (D_{\mu} \Phi)^{\dag} D_{\nu} \Phi -  {1 \over 
16 \pi e^2} F_{\mu\nu}F^{\mu\nu} \bigr],
\cr
\sp(3.0)} \eqno(1)$$
%-----------------------------------------------------------------------
where $\Phi$ is a complex scalar field,  
%%%%%%%%%%%%%%%%%%%%%%%%%%%%Equation%%%%%%%%%%%%%%%%%%%%%%%%%%%%%%%%%%%
$$ \eqalign{ \sp(2.0)
D_{\mu} \Phi = \partial_{\mu}\Phi + i A_{\mu} \Phi
\cr
\sp(3.0)} \eqno(2)$$
%-----------------------------------------------------------------------
is its covariant derivative, $A_{\mu}$ is the electromagnetic field, 
$F_{\mu\nu}$ is the corresponding field strength as usual given by
%%%%%%%%%%%%%%%%%%%%%%%%%%%%Equation%%%%%%%%%%%%%%%%%%%%%%%%%%%%%%%%%%%
$$ \eqalign{ \sp(2.0)
F_{\mu\nu} = \partial_{\mu} A_{\nu} - \partial_{\nu} A_{\mu},
\cr
\sp(3.0)} \eqno(3)$$
%-----------------------------------------------------------------------
and $e$ is the electric charge of $\Phi$. To clarify the four dimensional 
meaning we put the suffix $(4)$ in front of the metric tensor and the 
curvature scalar. We follow the conventions adopted in the MTW textbook 
[3] and use the natural units $G = \hbar = c = 1$. The Greek letters 
$\mu, \nu, ...$ take 0, 1, 2, and 3, on the other hand, the Latin ones $a, b, 
...$ do 0 and 1. Of course, the inclusion of other matter fields and the 
cosmological constants in this formalism is straightforward even if we 
limit ourselves to the action (1).

The most general spherically symmetric assumption for the metric is
%%%%%%%%%%%%%%%%%%%%%%%%%%%%Equation%%%%%%%%%%%%%%%%%%%%%%%%%%%%%%%%%%%
$$ \eqalign{ \sp(2.0)
ds^2 &= {}^{(4)}g_{\mu\nu} dx^{\mu} dx^{\nu},
\cr
     &= g_{ab} dx^a dx^b + \phi^2 ( d\theta^2 + \sin^2\theta d\varphi^2 ), 
\cr
\sp(3.0)} \eqno(4)$$
%-----------------------------------------------------------------------
where the two dimensional metric $g_{ab}$ and the radial function $\phi$ 
are the functions of only the two dimensional coordinates $x^a$. The 
substitution of (4) into (1) and then integration over the angular 
coordinates $(\theta, \varphi)$ leads to the following two dimensional 
effective action
%%%%%%%%%%%%%%%%%%%%%%%%%%%%Equation%%%%%%%%%%%%%%%%%%%%%%%%%%%%%%%%%%%
$$ \eqalign { \sp(2.0)
S &= {1 \over 2} \int \ d^2 x \sqrt{-g} \bigl[ 1 + g^{ab} \partial_a \phi 
\partial_b \phi +  {1 \over 2} R \phi^2 \bigr] 
\cr
&\qquad- \int \ d^2 x \sqrt{-g} \phi^2 g^{ab} (D_a 
\Phi)^{\dag} D_b \Phi - {1 \over 4} \int \ d^2 x \sqrt{-g} \phi^2 F_{ab} 
F^{ab},
\cr
\sp(3.0)} \eqno(5)$$
%-----------------------------------------------------------------------
where 
%%%%%%%%%%%%%%%%%%%%%%%%%%%%Equation%%%%%%%%%%%%%%%%%%%%%%%%%%%%%%%%%%%
$$ \eqalign{ \sp(2.0)
D_a \Phi = \partial_a \Phi + i e A_a \Phi.
\cr
\sp(3.0)} \eqno(6)$$
%-----------------------------------------------------------------------

Next let us rewrite the action (5) into the first-order ADM form. As 
mentioned before, we shall take the $x^1$ coordinate as time 
to cover the internal region of a black hole by spacelike hypersurfaces. 
The appropriate ADM splitting of (1+1)-dimensional spacetime is given by 
%%%%%%%%%%%%%%%%%%%%%%%%%%%%Equation%%%%%%%%%%%%%%%%%%%%%%%%%%%%%%%%%%%
$$ \eqalign{ \sp(2.0)
g_{ab} = \left(\matrix{ \gamma  & \alpha \cr
              \alpha & {\alpha^2 \over \gamma} - \beta^2 \cr} \right),
\cr
\sp(3.0)} \eqno(7)$$
%-----------------------------------------------------------------------
and the normal unit vector $n^a$ which is orthogonal to the hypersurface 
$x^1 = const$ reads 
%%%%%%%%%%%%%%%%%%%%%%%%%%%%Equation%%%%%%%%%%%%%%%%%%%%%%%%%%%%%%%%%%%
$$ \eqalign{ \sp(2.0)
n^a = ({\alpha \over {\beta \gamma}}, - {1 \over \beta}).
\cr
\sp(3.0)} \eqno(8)$$
%-----------------------------------------------------------------------
The induced metric on the hypersurfaces, that is, the projection operator 
over $x^1 = const$ hypersurfaces, is given by 
%%%%%%%%%%%%%%%%%%%%%%%%%%%%Equation%%%%%%%%%%%%%%%%%%%%%%%%%%%%%%%%%%%
$$ \eqalign{ \sp(2.0)
h^{ab} = g^{ab} + n^a n^b.
\cr
\sp(3.0)} \eqno(9)$$
%-----------------------------------------------------------------------
It is easy to check that $h^{ab}$ is indeed the projection operator by 
substituting (7) and (8) into (9).

The extrinsic curvature $K_{ab}$, its trace $K$, and the scalar curvature 
$R$ are 
given by [15]
%%%%%%%%%%%%%%%%%%%%%%%%%%%%Equation%%%%%%%%%%%%%%%%%%%%%%%%%%%%%%%%%%%
$$ \eqalign{ \sp(2.0)
K_{ab} = K_{ba} = h_a ^c \nabla_c n_b , 
\cr
\sp(3.0)} \eqno(10)$$
%-----------------------------------------------------------------------
%%%%%%%%%%%%%%%%%%%%%%%%%%%%Equation%%%%%%%%%%%%%%%%%%%%%%%%%%%%%%%%%%%
$$ \eqalign{ \sp(2.0)
K = g^{ab} K_{ab} = \nabla_a n^a = {1 \over \sqrt{-g}} \partial_a 
(\sqrt{-g} n^a) , 
\cr
\sp(3.0)} \eqno(11)$$
%-----------------------------------------------------------------------
and
%%%%%%%%%%%%%%%%%%%%%%%%%%%%Equation%%%%%%%%%%%%%%%%%%%%%%%%%%%%%%%%%%%
$$ \eqalign{ \sp(2.0)
R = 2 n^a \partial_a K + 2K^2 - 2 \nabla_c (n^a \nabla_a n^c).
\cr
\sp(3.0)} \eqno(12)$$
%-----------------------------------------------------------------------
A straightforward calculation yields
%%%%%%%%%%%%%%%%%%%%%%%%%%%%Equation%%%%%%%%%%%%%%%%%%%%%%%%%%%%%%%%%%%
$$ \eqalign{ \sp(2.0)
K = - {\gamma^\prime \over {2\beta\gamma}} + {\dot \alpha \over 
{\beta\gamma}} - {\alpha \over {2\beta\gamma^2}} \dot \gamma,
\cr
\sp(3.0)} \eqno(13)$$
%-----------------------------------------------------------------------
and 
%%%%%%%%%%%%%%%%%%%%%%%%%%%%Equation%%%%%%%%%%%%%%%%%%%%%%%%%%%%%%%%%%%
$$ \eqalign{ \sp(2.0)
R = 2 n^a \partial_a K + 2K^2 - {2 \over {\beta\sqrt{\gamma}}} \partial_0 
({\dot \beta \over {\sqrt{\gamma}}}),
\cr
\sp(3.0)} \eqno(14)$$
%-----------------------------------------------------------------------
where ${\partial \over {\partial x^0}} = \partial_0$ and ${\partial \over 
{\partial x^1}} = \partial_1$ are also denoted by an overdot and a prime, 
respectively. As a result, the action (5) can be written as
%%%%%%%%%%%%%%%%%%%%%%%%%%%%Equation%%%%%%%%%%%%%%%%%%%%%%%%%%%%%%%%%%%
$$ \eqalign{ \sp(2.0)
S &= \int d^2x L  =\int d^2x \bigl[ {1 \over 2} \beta \sqrt{\gamma} 
\bigl\{ 1 - (n^a \partial_a 
 \phi)^2 + {1 \over \gamma} \dot \phi^2 - K n^a \partial_a (\phi^2) 
\cr
&\qquad+ {\dot \beta \over {\beta\gamma}} \partial_0 (\phi^2) \bigr\} 
+ \beta \sqrt{\gamma} \phi^2 \bigl\{{ | n^a D_a \Phi |^2 - {1 \over \gamma}
|D_0 \Phi |^2 }\bigr\} + {1 \over 2} \beta \sqrt{\gamma} \phi^2 E^2 \bigr] 
\cr
&\qquad+ \int d^2x \bigl[ {1 \over 2} \partial_a 
 (\beta \sqrt{\gamma} K n^a \phi^2 ) 
- {1 \over 2} \partial_0 ({\dot \beta 
 \over {\sqrt{\gamma}} \phi^2}) \bigr]\hfill,
\cr
\sp(3.0)} \eqno(15)$$

%-----------------------------------------------------------------------
where
%%%%%%%%%%%%%%%%%%%%%%%%%%%%Equation%%%%%%%%%%%%%%%%%%%%%%%%%%%%%%%%%%%
$$ \eqalign{ \sp(2.0)
E&= {1 \over \sqrt{-g}} F_{01},
\cr
 &= {1 \over \beta \sqrt{\gamma}} (\dot A_1 - A_0^{\prime}).
\cr
\sp(3.0)} \eqno(16)$$
%-----------------------------------------------------------------------

By differentiating the action (15) with respect to the canonical 
variables $\Phi (\Phi^{\dag}), \phi, \gamma_0$ and $A_0$ we have the 
corresponding conjugate momenta $p_{\Phi} ({p_{\Phi^{\dag}}}), p_{\phi}, 
p_{\gamma}$ and $p_A$ 
%%%%%%%%%%%%%%%%%%%%%%%%%%%%Equation%%%%%%%%%%%%%%%%%%%%%%%%%%%%%%%%%%%
$$ \eqalign{ \sp(2.0)
p_{\Phi} = - \sqrt{\gamma} \phi^2 (n^a D_a\Phi)^{\dag},
\cr
\sp(3.0)} \eqno(17)$$
%-----------------------------------------------------------------------
%%%%%%%%%%%%%%%%%%%%%%%%%%%%Equation%%%%%%%%%%%%%%%%%%%%%%%%%%%%%%%%%%%
$$ \eqalign{ \sp(2.0)
p_{\phi} = \sqrt{\gamma} n^a \partial_a \phi + \sqrt{\gamma} K \phi,
\cr
\sp(3.0)} \eqno(18)$$
%-----------------------------------------------------------------------
%%%%%%%%%%%%%%%%%%%%%%%%%%%%Equation%%%%%%%%%%%%%%%%%%%%%%%%%%%%%%%%%%%
$$ \eqalign{ \sp(2.0)
p_{\gamma} = {1 \over 4 \sqrt \gamma} n^a \partial_a (\phi^2),
\cr
\sp(3.0)} \eqno(19)$$
%-----------------------------------------------------------------------
%%%%%%%%%%%%%%%%%%%%%%%%%%%%Equation%%%%%%%%%%%%%%%%%%%%%%%%%%%%%%%%%%%
$$ \eqalign{ \sp(2.0)
p_A = - \phi^2 E.
\cr
\sp(3.0)} \eqno(20)$$
%-----------------------------------------------------------------------
The Hamiltonian $H$ which is defined as
%%%%%%%%%%%%%%%%%%%%%%%%%%%%Equation%%%%%%%%%%%%%%%%%%%%%%%%%%%%%%%%%%%
$$ \eqalign{ \sp(2.0)
H = \int dx^0 ( p_{\Phi} \Phi^{\prime} + {p_{\Phi^{\dag}}} 
\Phi^{\prime \dag} 
+ p_{\phi} \phi^{\prime} + p_{\gamma} \gamma^{\prime} + p_A A_0^{\prime} - L ) 
\cr
\sp(3.0)} \eqno(21)$$
%-----------------------------------------------------------------------
is expressed by a linear combination of constraints as expected 
%%%%%%%%%%%%%%%%%%%%%%%%%%%%Equation%%%%%%%%%%%%%%%%%%%%%%%%%%%%%%%%%%%
$$ \eqalign{ \sp(2.0)
H = \int dx^0 ( \alpha H_0 + \beta H_1 + A_1 H_2 ), 
\cr
\sp(3.0)} \eqno(22)$$
%-----------------------------------------------------------------------
where
%%%%%%%%%%%%%%%%%%%%%%%%%%%%Equation%%%%%%%%%%%%%%%%%%%%%%%%%%%%%%%%%%%
$$ \eqalign{ \sp(2.0)
H_0 = {1 \over \gamma} [p_\Phi D_0 \Phi + {p_{\Phi^{\dag}}} (D_0 \Phi)^{\dag}]
 + {1 \over \gamma} p_\phi \dot \phi - 2 \dot p_\gamma - {1 \over 
\gamma} p_\gamma \dot \gamma, 
\cr
\sp(3.0)} \eqno(23)$$
%-----------------------------------------------------------------------
%%%%%%%%%%%%%%%%%%%%%%%%%%%%Equation%%%%%%%%%%%%%%%%%%%%%%%%%%%%%%%%%%%
$$ \eqalign{ \sp(2.0)
H_1 &= {1 \over {\sqrt{\gamma} \phi^2}} p_{\Phi} {p_{\Phi^{\dag}}} - 
{\sqrt {\gamma} \over 2} - {\dot \phi^2 \over {2 \sqrt{\gamma}}} 
+ \partial_0 ( {\partial_0 (\phi^2) \over {2 \sqrt{\gamma}}}) 
\cr
&\qquad+ {\phi^2 \over \sqrt{\gamma}} | D_0 \Phi |^2 - {2 \sqrt{\gamma} 
\over \phi} p_\phi p_\gamma + {2 \gamma \sqrt{\gamma} \over \phi^2} 
p_\gamma ^2  + {\sqrt{\gamma} \over {2 \phi^2}} p_A ^2 , 
\cr
\sp(3.0)} \eqno(24)$$
%-----------------------------------------------------------------------
%%%%%%%%%%%%%%%%%%%%%%%%%%%%Equation%%%%%%%%%%%%%%%%%%%%%%%%%%%%%%%%%%%
$$ \eqalign{ \sp(2.0)
H_2 = - ie (p_{\Phi} \Phi - p_{\Phi^{\dag}} \Phi^{\dag}) - \dot p_A.
\cr
\sp(3.0)} \eqno(25)$$
%-----------------------------------------------------------------------
Note that $\alpha$, $\beta$ and $A_1$ are non-dynamical Lagrange 
multiplier fields.

The action can be cast into the first-order ADM canonical form by the 
dual Legendre transformation 
%%%%%%%%%%%%%%%%%%%%%%%%%%%%Equation%%%%%%%%%%%%%%%%%%%%%%%%%%%%%%%%%%%
$$ \eqalign{ \sp(2.0)
S = \int d x^1 \bigl[ \int d x^0 ( p_{\Phi} \Phi^{\prime} + 
{p_{\Phi^{\dag}}} \Phi^{\prime \dag} 
+ p_{\phi} \phi^{\prime} + p_{\gamma} {\gamma}^{\prime} + p_A 
A_0^{\prime} ) - H \bigr]. 
\cr
\sp(3.0)} \eqno(26)$$
%-----------------------------------------------------------------------
As Regge and Teitelboim pointed out [16], in order to have the correct 
Hamiltonian which produces the Einstein equations through the Hamilton 
equations, one has to supplement the surface terms to the Hamiltonian 
(22). In the present formalism, since we take the variation of  all the
fields to be zero at both the singularity and the apparent horizon we 
do not have to add any surface terms to the Hamiltonian.

\vskip 12pt
\leftline{\bf 3. Black hole radiation}	
\par
We now turn our attention to an application of the canonical formalism 
constructed in the previous section for understanding the Hawking 
radiation [12] from the viewpoint in the internal region of a black hole 
in quantum gravity. A similar analysis was performed in the external 
region of a black hole by Tomimatsu [13]. 

To consider the simplest model of the Hawking radiation, let us turn off 
the electromagnetic field and treat the neutral scalar field by which the 
constraint $H_2$ generating the $U(1)$ gauge transformations identically 
vanishes. Moreover, we shall use the ingoing Vaidya metric to express 
the black hole radiation. The treatment of the case of the outgoing 
Vaidya metric can be made in a perfectly similar way. Here it is useful 
to introduce the advanced time coordinate $v$ defined as [3]
%%%%%%%%%%%%%%%%%%%%%%%%%%%%Equation%%%%%%%%%%%%%%%%%%%%%%%%%%%%%%%%%%%
$$ \eqalign{ \sp(2.0)
v = t + r^*,
\cr
\sp(3.0)} \eqno(27)$$
%-----------------------------------------------------------------------
with the tortoise coordinate
%%%%%%%%%%%%%%%%%%%%%%%%%%%%Equation%%%%%%%%%%%%%%%%%%%%%%%%%%%%%%%%%%%
$$ \eqalign{ \sp(2.0)
r^* = \int^r {dr \over {1 - {2M \over r}}},
\cr
\sp(3.0)} \eqno(28)$$
%-----------------------------------------------------------------------
where $M = M(x^a)$ is the mass function. Here let us define the two 
dimensional coordinate $x^a$ by 
%%%%%%%%%%%%%%%%%%%%%%%%%%%%Equation%%%%%%%%%%%%%%%%%%%%%%%%%%%%%%%%%%%
$$ \eqalign{ \sp(2.0)
x^a = (x^0, x^1) = (v - r, r).
\cr
\sp(3.0)} \eqno(29)$$
%-----------------------------------------------------------------------
Note that we have chosen $x^1 = r$ as mentioned above. And we fix the 
gauge freedoms corresponding to the two dimensional 
reparametrization invariances by the gauge conditions
%%%%%%%%%%%%%%%%%%%%%%%%%%%%Equation%%%%%%%%%%%%%%%%%%%%%%%%%%%%%%%%%%%
$$ \eqalign{ \sp(2.0)
g_{ab} &= \left(\matrix{ \gamma  & \alpha \cr
              \alpha & {\alpha^2 \over \gamma} - \beta^2 \cr} \right),
\cr
 &= \left(\matrix{ -(1 - {2M \over r})   &  {2M \over r} \cr
              {2M \over r} & 1 + {2M \over r}} \right).
\cr
\sp(3.0)} \eqno(30)$$
%-----------------------------------------------------------------------
From these equations the two dimensional line element takes a form of the 
Vaidya metric
%%%%%%%%%%%%%%%%%%%%%%%%%%%%Equation%%%%%%%%%%%%%%%%%%%%%%%%%%%%%%%%%%%
$$ \eqalign{ \sp(2.0)
ds^2 &= g_{ab} dx^a dx^b,
\cr
     &= -(1 - {2M \over r}) dv^2 + 2 dv dr.
\cr
\sp(3.0)} \eqno(31)$$
%-----------------------------------------------------------------------
This is , of course, rewritten in terms of the $(t, r)$ coordinates by using 
Eq.s (27), (28) into the Schwarzschild form 
%%%%%%%%%%%%%%%%%%%%%%%%%%%%Equation%%%%%%%%%%%%%%%%%%%%%%%%%%%%%%%%%%%
$$ \eqalign{ \sp(2.0)
ds^2 = -(1 - {2M \over r}) dt^2 + {1 \over {(1 - {2M \over r})}} dr^2.
\cr
\sp(3.0)} \eqno(32)$$
%-----------------------------------------------------------------------
For a dynamical black hole, it is useful to consider the local definition 
of horizon, i.e., the apparent horizon, rather than the global one, the 
event horizon. The apparent horizon is 
then defined as
%%%%%%%%%%%%%%%%%%%%%%%%%%%%Equation%%%%%%%%%%%%%%%%%%%%%%%%%%%%%%%%%%%
$$ \eqalign{ \sp(2.0)
x^1 = r = 2 M (x^0, x^1).
\cr
\sp(3.0)} \eqno(33)$$
%-----------------------------------------------------------------------

Following the idea in the reference [13], let us attempt to solve the 
Hamiltonian and supermomentum constraints only in the vicinity of the 
apparent horizon. Near the apparent horizon, by intuition we would be 
able to make an approximation 
%%%%%%%%%%%%%%%%%%%%%%%%%%%%Equation%%%%%%%%%%%%%%%%%%%%%%%%%%%%%%%%%%%
$$ \eqalign{ \sp(2.0)
\Phi \approx \Phi(v), M \approx M(v), \phi \approx r.
\cr
\sp(3.0)} \eqno(34)$$
%-----------------------------------------------------------------------
From now on we shall use $\approx$ to indicate the equalities which hold 
approximately near the apparent horizon. Indeed one can prove the above 
assumptions (34) to be consistent with the field equations as follows.

The field equations stemming from the action (5) are given by
%%%%%%%%%%%%%%%%%%%%%%%%%%%%Equation%%%%%%%%%%%%%%%%%%%%%%%%%%%%%%%%%%%
$$ \eqalign{ \sp(2.0)
- {2 \over \phi} \nabla_a \nabla_b \phi + {2 \over \phi} g_{ab} \nabla_c 
\nabla^c \phi + {1 \over \phi^2} g_{ab} \partial_c \phi \partial^c \phi 
- {1 \over \phi^2} g_{ab} = 2 (\partial_a \Phi \partial_b \Phi - {1 \over 2} 
g_{ab} \partial_c \Phi \partial^c \Phi),
\cr
\sp(3.0)} \eqno(35)$$
%-----------------------------------------------------------------------
%%%%%%%%%%%%%%%%%%%%%%%%%%%%Equation%%%%%%%%%%%%%%%%%%%%%%%%%%%%%%%%%%%
$$ \eqalign{ \sp(2.0)
{1 \over \sqrt{-g}} \partial_a (\sqrt{-g} g^{ab} \partial_b \phi) - {1 
\over 2} R \phi = - \phi \partial_a \Phi \partial^a \Phi,
\cr
\sp(3.0)} \eqno(36)$$
%-----------------------------------------------------------------------
%%%%%%%%%%%%%%%%%%%%%%%%%%%%Equation%%%%%%%%%%%%%%%%%%%%%%%%%%%%%%%%%%%
$$ \eqalign{ \sp(2.0)
\partial_a (\sqrt{-g} \phi^2 g^{ab} \partial_b \Phi) = 0.
\cr
\sp(3.0)} \eqno(37)$$
%-----------------------------------------------------------------------
As before the Vaidya metric is given by 
%%%%%%%%%%%%%%%%%%%%%%%%%%%%Equation%%%%%%%%%%%%%%%%%%%%%%%%%%%%%%%%%%%
$$ \eqalign{ \sp(2.0)
g_{ab} &= \left(\matrix{ g_{vv}  & g_{vr} \cr
              g_{rv} & g_{rr} \cr} \right),
\cr
       &= \left(\matrix{ - (1 - {2M \over r})  & +1 \cr
              +1 & 0 \cr} \right).
\cr
\sp(3.0)} \eqno(38)$$
%-----------------------------------------------------------------------
To prove Eq.(34), in the vicinity of the apparent horizon we make an 
ansatz that the fields have a form 
%%%%%%%%%%%%%%%%%%%%%%%%%%%%Equation%%%%%%%%%%%%%%%%%%%%%%%%%%%%%%%%%%%
$$ \eqalign{ \sp(2.0)
M \approx M(v), \phi \approx r,
\cr
\sp(3.0)} \eqno(39)$$
%-----------------------------------------------------------------------
but the scalar field is still given by $\Phi \approx \Phi(v,r)$. Under 
this ansatz, the field equations (35), (36), and (37) become
%%%%%%%%%%%%%%%%%%%%%%%%%%%%Equation%%%%%%%%%%%%%%%%%%%%%%%%%%%%%%%%%%%
$$ \eqalign{ \sp(2.0)
\partial_r \Phi \approx 0,
\cr
\sp(3.0)} \eqno(40)$$
%-----------------------------------------------------------------------
%%%%%%%%%%%%%%%%%%%%%%%%%%%%Equation%%%%%%%%%%%%%%%%%%%%%%%%%%%%%%%%%%%
$$ \eqalign{ \sp(2.0)
\partial_v \Phi \approx {\sqrt{\partial_v M} \over r} \approx 
{\sqrt{\partial_v M} \over 2M},
\cr
\sp(3.0)} \eqno(41)$$
%-----------------------------------------------------------------------
%%%%%%%%%%%%%%%%%%%%%%%%%%%%Equation%%%%%%%%%%%%%%%%%%%%%%%%%%%%%%%%%%%
$$ \eqalign{ \sp(2.0)
\partial_r \partial_v \Phi \approx \partial_v \partial_r \Phi
\approx -{\sqrt{\partial_v M} \over r^2} \approx -{\sqrt{\partial_v M} 
\over 4M^2}.
\cr
\sp(3.0)} \eqno(42)$$
%-----------------------------------------------------------------------
Then one can easily find the solution which satisfies at the same time 
Eqs. (40), (41), and (42) 
%%%%%%%%%%%%%%%%%%%%%%%%%%%%Equation%%%%%%%%%%%%%%%%%%%%%%%%%%%%%%%%%%%
$$ \eqalign{ \sp(2.0)
\Phi(v, r) = ({1 - {2M \over r}})^2 {1 \over {4 \sqrt{\partial_v M}}} + 
\int^v dv {\sqrt{\partial_v M} \over {2 M(v)}}.
\cr
\sp(3.0)} \eqno(43)$$
%-----------------------------------------------------------------------
Consequently, 
%%%%%%%%%%%%%%%%%%%%%%%%%%%%Equation%%%%%%%%%%%%%%%%%%%%%%%%%%%%%%%%%%%
$$ \eqalign{ \sp(2.0)
\Phi(v, r) \approx  \int^v dv {\sqrt{\partial_v M} \over {2 M(v)}},
\cr
\sp(3.0)} \eqno(44)$$
%-----------------------------------------------------------------------
which means that one can set $\Phi(v, r) \approx \Phi(v)$ like other 
fields in the vicinity of the apparent horizon. Thus our assumtions (34) 
are consistent with fields equations.

Near the apparent horizon (33), Eq.(30) yields
%%%%%%%%%%%%%%%%%%%%%%%%%%%%Equation%%%%%%%%%%%%%%%%%%%%%%%%%%%%%%%%%%%
$$ \eqalign{ \sp(2.0)
\alpha \approx +1 , \gamma = {1 \over \beta^2} \approx 0,
\cr
\sp(3.0)} \eqno(45)$$
%-----------------------------------------------------------------------
and the canonical conjugate momenta (17), (18) and (19) are given 
approximately as
%%%%%%%%%%%%%%%%%%%%%%%%%%%%Equation%%%%%%%%%%%%%%%%%%%%%%%%%%%%%%%%%%%
$$ \eqalign{ \sp(2.0)
p_{\Phi} \approx - \phi^2 \partial_v \Phi,
\cr
p_{\phi} \approx - {1 \over \gamma} \partial_v M,
\cr
p_{\gamma} \approx -{1 \over 2} \phi \approx -M.
\cr
\sp(3.0)} \eqno(46)$$
%-----------------------------------------------------------------------
Moreover, the two constraints are proportional to each other
%%%%%%%%%%%%%%%%%%%%%%%%%%%%Equation%%%%%%%%%%%%%%%%%%%%%%%%%%%%%%%%%%%
$$ \eqalign{ \sp(2.0)
- \gamma H_0 &\approx \sqrt{\gamma} H_1,
\cr
             &\approx {1 \over \phi^2} p_{\Phi}^2 + \gamma p_\phi. 
\cr
\sp(3.0)} \eqno(47)$$
%-----------------------------------------------------------------------
Here we have one important remark. Just at the apparent horizon $\gamma$ 
becomes zero so that the various equalities approximately hold when
 we restrict our attention to the interior region 
near but not at the apparent horizon. Under this sort of regularization 
$\gamma$ takes a small but finite value. 

An imposition of the constraint (47) as an operator equation on the state 
produces the Wheeler-DeWitt equation
%%%%%%%%%%%%%%%%%%%%%%%%%%%%Equation%%%%%%%%%%%%%%%%%%%%%%%%%%%%%%%%%%%
$$ \eqalign{ \sp(2.0)
i {\partial \Psi \over {\partial \phi}} = - {1 \over {\gamma 
\phi^2}} {\partial^2 \over {\partial \Phi^2}} \Psi.
\cr
\sp(3.0)} \eqno(48)$$
%-----------------------------------------------------------------------
This Wheeler-DeWitt equation can be interpreted as the Schr$\ddot o$dinger 
equation with the superspace Hamiltonian $H_S = 
{1 \over {\gamma \phi^2}} p_{\Phi}^2$.  

Now it is easy to find a special solution of the Wheeler-DeWitt 
equation (48) by the method of separation of variables. The result is 
%%%%%%%%%%%%%%%%%%%%%%%%%%%%Equation%%%%%%%%%%%%%%%%%%%%%%%%%%%%%%%%%%%
$$ \eqalign{ \sp(2.0)
\Psi = (B e^{\sqrt A \Phi} + C e^{-\sqrt A \Phi} ) e^{- i{A \over {\gamma 
\phi}} },
\cr
\sp(3.0)} \eqno(49)$$
%-----------------------------------------------------------------------
where $A$, $B$, and $C$ are integration constants. If one defines an 
expectation value $< \cal O >$ of an operator $\cal O$ as
%%%%%%%%%%%%%%%%%%%%%%%%%%%%Equation%%%%%%%%%%%%%%%%%%%%%%%%%%%%%%%%%%%
$$ \eqalign{ \sp(2.0)
< {\cal O} > = {1 \over {\int d\Phi |\Psi|^2 }} \int d\Phi \Psi^* {\cal O} 
\Psi,
\cr
\sp(3.0)} \eqno(50)$$
%-----------------------------------------------------------------------
one can calculate $< \partial_v M >$ by using either (46) or (47)
%%%%%%%%%%%%%%%%%%%%%%%%%%%Equation%%%%%%%%%%%%%%%%%%%%%%%%%%%%%%%%%%%
$$ \eqalign{ \sp(2.0)
< \partial_v M > = - {A \over {\phi^2}}.
\cr
\sp(3.0)} \eqno(51)$$
%-----------------------------------------------------------------------
This equation shows the black hole radiation when one chooses the 
constant $A$ to be a positive constant $k^2$. Then the mass-loss rate 
becomes
%%%%%%%%%%%%%%%%%%%%%%%%%%%%Equation%%%%%%%%%%%%%%%%%%%%%%%%%%%%%%%%%%%
$$ \eqalign{ \sp(2.0)
< \partial_v M > = - {k^2 \over {4 < M >^2}}.
\cr
\sp(3.0)} \eqno(52)$$
%-----------------------------------------------------------------------
This result exactly corresponds to that calculated by Hawking in the 
semiclassical approach [12] and by Tomimatsu in the exterior region of 
the apparent horizon in quantum gravity [13]. Thus a black hole 
completely evaporates 
within a finite time. However, it is worth pointing out the difference 
between the Howking formulation and the present one. In the Hawking's 
semiclassical approach the gravitational field is fixed as a classical 
background and only the matter field is treated to be quantum-mechanical. 
By contrast, our present formulation is purely quantum-mechanical.

At this stage, we must pay our attention to the boundary condition 
for the physical state. Our physical state which satisfies the 
Wheeler-Dewitt equation 
%%%%%%%%%%%%%%%%%%%%%%%%%%%%Equation%%%%%%%%%%%%%%%%%%%%%%%%%%%%%%%%%%%
$$ \eqalign{ \sp(2.0)
\Psi = (B e^{|k| \Phi} + C e^{-|k| \Phi} ) e^{- i{k^2 \over 
{\gamma \phi}}} ,
\cr
\sp(3.0)} \eqno(53)$$
%-----------------------------------------------------------------------
certainly does neither satisfy the Dirichlet boundary condition $\Psi 
\rightarrow 0$ for $|\Phi| \rightarrow \infty$ nor its norm 
${\int d\Phi |\Psi|^2 }$ is finite. Of course, these requirements might be 
too heavy since we do not have any physical principle to choose the 
approriate boundary condition and the present formalism is too restrictive
to select the correct definition of the inner product. Nevertheless, it 
seems interesting to impose these requirements on our physical 
state (53). Although we have derived the expression (53) which is valid 
only in the vicinity of 
the apparent horizon, we would expect that the qualitative features 
remain valid in the whole interior region of a black hole except  
terms coming from some higher-order quantum corrections. Near the 
spacetime singularity  the matter field $\Phi$ would strongly fluctuate 
owing to huge quantum effects, thus the Dirichlet boundary condition 
seems to be most suitable to supress this unwieldy behavior of the 
state. It seems that one of candidates satisfying the Dirichlet boundary 
condition would be of the form
%%%%%%%%%%%%%%%%%%%%%%%%%%%%Equation%%%%%%%%%%%%%%%%%%%%%%%%%%%%%%%%%%%
$$ \eqalign{ \sp(2.0)
\Psi = (B e^{|k| \Phi} + C e^{-|k| \Phi} ) e^{- K \Phi^2} e^{- i{k^2 \over 
{\gamma \phi}}} ,
\cr
\sp(3.0)} \eqno(54)$$
%-----------------------------------------------------------------------
where $K$ is a certain positive constant. It is interesting that this 
state (54) also leads to the finite norm under the naively defined measure.
It would be wonderful if we can derive such a welcoming physical state by 
improving the present model in the future.

\vskip 12pt
\leftline{\bf 4. Conclusion}	
\par
In this article, we have constructed a canonical formalism which would be 
suitable for a description of a black hole inside the apparent horizon. 
The aim in mind to use this formalism is to give a basis for analysing 
various problems associated with an internal structure of a black hole in 
quantum gravity. As a 
concrete application based on this canonical formalism, we have studied 
the Hawking radiation from quantum mechanical viewpoint and shown that 
the physical state satisfying the Wheeler-DeWitt equation gives us the 
same mass-loss rate as Hawking's semiclassical calculation. Thus the 
canonical formalism 
constructed in this paper seems to give us to a framework in 
understanding the quantum mechanical properties of a black hole.

In future works, we would like to investigate further other recently 
developed problems such as the mass inflation, the smearing of the black 
hole singularities, and the strong cosmic censorship by the 
present formalism. These studies are currently under active 
investigation, hence the results will appear in the future publications 
[17].

\vskip 12pt
\leftline{\bf References}
\centerline{ } %
\par
\item{[1]} P.A.M.Dirac, Lectures on Quantum Mechanics (Yeshiva University, 1964). 

\item{[2]} P.Arnowitt, S.Deser, and C.W.Misner, in Gravitation: An 
Introduction to Current Research, edited by L.Witten (Wiley, New York, 
1962).

\item{[3]} C.W.Misner, K.S.Thorne, and J.A.Wheeler, Gravitation (Freeman, 1973). 

\item{[4]} P.Thomi, B.Isaak, and P.Hajicek, Phys. Rev. {\bf D30} (1984) 
1168.

\item{[5]} P.Hajicek, Phys. Rev. {\bf D30} (1984) 1178.

\item{[6]} K.V.Kuchar, Phys. Rev. {\bf D50} (1994) 3961.

\item{[7]} E.Poisson and W.Israel, Phys. Rev. {\bf D41} (1990) 1796.

\item{[8]} A.Ori, Phys. Rev. Lett. {\bf 67} (1991) 789; {\bf 68} (1992) 
2117.

\item{[9]} P.R.Brady and C.M.Chambers, Phys. Rev. {\bf D51} (1995) 4177; 
P.R.Brady and J.D.Smith, Phys. Rev. Lett. {\bf 75} (1995) 1256.

\item{[10]} A.Hosoya, Class. Quant. Grav. {\bf 12} (1995) 2967.

\item{[11]} A.Hosoya and I.Oda, ``Black Hole Singularity and Generalized 
Quantum Affine Parameter'', TIT/HEP-334/COSMO-73, EDO-EP-5, gr-qc/9605069.

\item{[12]} S.W.Hawking, Comm. Math. Phys. {\bf 43} (1975) 199.

\item{[13]} A.Tomimatsu, Phys. Lett. {\bf B289} (1992) 283.

\item{[14]} N.D.Birrell and P.C.W.Davies, Quantum Fields in Curved Space 
(Cambridge University Press, Cambridge, 1982).

\item{[15]} R.M.Wald, General Relativity (The University of Chicago Press, 
1984).

\item{[16]} T.Regge and C.Teitelboim, Ann. Phys. (N.Y.) {\bf 88} (1974) 286.

\item{[17]} A.Hosoya and I.Oda, ``Mass Inflation in Quantum Gravity'', to 
appear

\endpage
%

%=======================================================================%
%
\bye